\documentclass[aps,twocolumn,prl,superscriptaddress,amsmath,showpacs,tightenlines]{revtex4}
\usepackage{graphicx}
\usepackage{epsfig}
\usepackage{color}
\usepackage[normalem]{ulem}

\begin{document}

\title{Effects of interface electric field on the magnetoresistance in spin devices}

\author{T. Tanamoto, 
        M. Ishikawa,
        T. Inokuchi,
        H. Sugiyama, 
        and Y.Saito}

\affiliation{Advanced LSI Technology Laboratory
	Corporate Research and Development Center,
	Toshiba Corporation
	1, Komukai Toshiba-cho, Saiwai-ku,
	Kawasaki 212-8582, Japan.}

\date{\today}

\begin{abstract}
An extension of the standard spin diffusion theory is presented by introducing 
a density-gradient (DG) term that is suitable for describing interface quantum tunneling phenomena.
The magnetoresistance (MR) ratio is modified by the DG term through an 
interface electric field. 
We have also carried out spin injection and detection measurements using four-terminal Si devices. 
The local measurement shows that the MR ratio changes depending on the current direction.
We show that the change of the MR ratio depending on the current direction 
comes from the DG term regarding the asymmetry of the two interface electronic structures. 
\end{abstract}

\maketitle

Spin injection and detection between silicon and magnetic material via tunneling barriers constitute 
one of the most important issues in spintronics, 
and the key factor that determines the performance of 
spin devices such as spin 
transistors~\cite{Tanaka,Appelbaum,Dash,Crowell,Suzuki,Li,Uemura,Jeon,Ando,Jansen,Saito1,InokuchiJAP,Ishikawa,Tanamoto,Zhang,TanaRT,Miao}.
The difference in electronic structure 
between ferromagnet (FM) and semiconductor (SC) generates  
the conductance mismatch between the two materials, 
which has inspired much important research regarding 
spin transport properties through the interface. 
With regard to the spin diffusion theory (standard theory)~\cite{ValetFert,Schmidt,Fert2001,Jaffres,Fukuma}, which has to a great content succeeded in 
explaining tunneling phenomena of nonmagnet (NM) sandwiched by two FMs, the challenge is 
to explain new phenomena that appears in SC:
Jansen {\it et al.} showed the effect of the depletion layer in the SC interface~\cite{Jansen0} 
and pointed out the importance of the consideration of
band structure at the interface.
Tran {\it et al.}\cite{Tran,Jansen1} discussed the relation between the localized states 
and the enhancement of spin accumulation signal at Co/Al${}_2$O${}_3$/GaAs interface. 
Yu {\it et al.}~\cite{Yu} derived the drift-diffusion equation without depletion layer
and 
Kameno {\it et al.}~\cite{Kameno} experimentally showed the contribution of 
the spin-drift current in the substrate.
Considering the usefulness of the standard spin diffusion theory and the fact that 
the effects of bulk properties such as spin drift can be taken into account by changing 
the lifetime or other parameters of the standard theory, 
it is desirable to extend the present standard theory so that it $directly$ includes interface effects 
that are specific to SC.

Here we theoretically extend the standard spin diffusion theory by 
taking into account the density-gradient (DG) theory that is 
derived from the quantum diffusion equation~\cite{Ancona1990}.
The DG theory can describe the interface phenomena such as quantum tunneling through SiO${}_2$
sandwiched by an electrode and Si substrate~\cite{Ancona2,Ancona3}.
This DG term appears when the electron density changes at the interface 
as a result of the band bending.
We apply this DG term which has been studied in the conventional silicon transistors 
to the two-spin-current model,
and we provide an analytic formula of the magnetoresistance (MR) ratio for FM/SC/FM structure.
We show that the DG term increases or decreases the MR ratio depending on the current direction
through FM/SC/FM structure 
when there is an $asymmetric$ electronic structure between the source interface and the drain interface. 
The direct way to realize the asymmetric electronic states at the two interfaces is 
to make the areas of the two electrodes different.
The different areas of the electrodes generate different depletion region depending 
on current direction.
We have carried out both {\it local} and {\it nonlocal} measurements for four-terminal silicon devices
that have asymmetric electrodes.
We show that experimentally obtained MR ratios
differ depending on the current direction (from source to drain or from drain to source).
The standard theory cannot explain this directionality of MR ratio, 
because the solution of the diffusion equation in the SC is symmetric between the source and the drain~\cite{tanaJJAP}.
The different electronic interfaces break this symmetric state in which the MR ratio has a maximum at 
the symmetric point.

{\it Formulation.}---
We take into account the DG term 
in the spin current to describe the quantum tunneling at the 
FM/SC interface through a tunneling barrier. The generalized chemical potential is described by
\begin{equation}
\mu_s = \mu_{s0} +2 b \left[ \frac{\nabla^2 \sqrt{n_s}}{\sqrt{n_s}} \right], 
\label{mu}
\end{equation}
where $\mu_{s0}$ is the conventional chemical potential ($s=\pm$). 
$n_s$ express a density of electron~\cite{Ancona1990,Ancona2,Ancona3}.
The second term express the DG term and 
$
b=\hbar^2/(2 mer_q)
$
is a coefficient of this term.
The parameter $r_q$ changes depending on the physical environment. 
Although Ancona {\it et al}. use $r_q=6 $ for high-temperature region, and $r_q=2$ for low-temperature region,
we take $r_q$ as a fitting parameter when we apply our theory to experiments. 
The generalized spin current is given by Ohm's law given by $
I_s=(\sigma_s/e) \partial  \mu_s/\partial z$, 
where $\sigma_s$ is a spin-dependent conductivity.

We assume that the same macroscopic diffusion equations as those 
of the standard theories~\cite{ValetFert,Fert2001,Schmidt}
hold:
$
\frac{e}{\sigma_\pm } \frac{\partial I_\pm}{\partial z}=\pm \frac{\mu_+-\mu_-}{l_\pm^2} 
$
and
$ 
 \frac{\partial^2 (\mu_+-\mu_-)}{\partial z^2} 
=\frac{(\mu_+-\mu_-)}{l_{sf}^2}  
$ 
with
$l_{sf}^{-2} =l_+^{-2}+l_-^{-2}$ as an average spin diffusion length. 
Combining with the current conservation relation given by $I_++I_-=I $(const), 
we have solutions for the differential equations in ferromagnetic region:
\begin{eqnarray}
\mu_{\pm}^\alpha (z) \!\!&\!=\!&\!\!(1\!-\!\beta^2) e\rho_F I_\alpha z
\nonumber \\ 
& + & K_1^\alpha \mp (1\! \mp \! \beta) \Delta \mu^\alpha  (z)  
\!+\psi_{\alpha b\pm}^F(z), \label{chem} \\
I_{\alpha \pm} (z) &=&(1\pm \beta) \frac{I_\alpha}{2} \mp \frac{1}{2e\rho_F l^F}  \Delta \nu^\alpha  (z)  
+I_{\alpha b\pm }^F(z) \label{curr},
\end{eqnarray}
with 
$\Delta \mu^\alpha  (z)=  K_m^\alpha  \sinh ([z-z_0^\alpha]/l^F)$,
and $\Delta \nu^\alpha  (z) =l_F \partial \Delta \mu^\alpha (z)/\partial z$
($z_0^\alpha$ is a center of $\alpha$-ferromagnet),
for both ferromagnetic electrodes $\alpha=L,R$.
$\rho_F$ is the resistivity of the FM. 
We obtained a similar formulation of the chemical potential $\mu_\pm(z)$ and the current $I_\pm(z)$ in the SC region. 
$K_m^\alpha$ and $K_1^\alpha$ are unknown coefficients
to be determined by the boundary conditions.
$\psi_{bL\pm}^\nu =  (2 b/[n_{\alpha\pm}^\nu]^{1/2}) \partial^2 [n_{\alpha\pm}^\nu]^{1/2}/\partial z^2$ 
with $\nu=F,N$ is the DG term (the second term in Eq.~(\ref{mu})), 
and $I_{bL\pm}^\nu$ is its derivative given by
$I_{b\alpha\pm}^\nu=  2  (\sigma_\pm /e)  \partial \psi_{b\alpha\pm}^\nu/\partial z$.


\begin{figure}
\centering
\includegraphics[width=8.3cm]{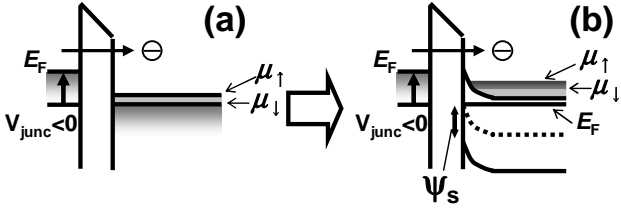}  
\caption{Energy diagram of magnetic contact with metal and n-type Si. (a) The tunnel junction between 
the left ferromagnet and the right metal, which is described by the 
standard theory. (b) Energy-band profile of a tunnel junction 
between ferromagnet and semiconductor in the depletion region. 
$\psi_s$ is a surface potential.
}
\label{place0}
\end{figure}


The same boundary conditions are applied to chemical potential
and current at $z=z_c$ in the standard theories~\cite{ValetFert,Fert2001}
are applied:
$\mu_\pm (z_c^+) -\mu_\pm^\alpha (z_c^-) =r_\pm^\alpha I_\pm (z_c)$
and $I_\pm (z_c^+) = I_\pm^\alpha (z_c^-)$. 
We also have a current conservation condition other than the interface regions such as
$I_L S_L=I S_N=I_R S_R$ where $S_\alpha$ and $S_{\rm N}$ are 
areas of $\alpha$-FM and the SC. 
We define
$r_\pm^\alpha = 2r_b^\alpha [1\pm \gamma^\alpha]$, 
$r_m^\alpha= [r_-^\alpha-r_+^\alpha]/4 =r_b^\alpha \gamma^\alpha$,
and $r_F=\rho_F l^F$, $r_N = \rho_N l^N$ where 
$\rho_N$ is the resistivity of the SC region.
The relations of the spin accumulation $\mu_{\alpha \pm}$ and the 
DG term can be derived from Eqs.(\ref{chem}) and (\ref{curr}),  and related equations given by
\begin{eqnarray}
& &  \Delta \mu (z_0) - \Delta \mu^L (z_0)   +\frac{r_p^L }{er_N}  \Delta \nu (z_0)  =r_m^LI_{L1}, 
\label{accum1}  \\
& & 
\frac{S_N}{er_N}  \Delta \nu (z_0) - \frac{S_L}{er_F}  \Delta \nu^L (z_0) =\beta S_N I_{L1}.  
\label{accum2}
\end{eqnarray}
Similar equations are obtained at the right interface depending on the parallel (P) and the antiparallel (AP) states.
Here $I_{\alpha 1}$  include the DG terms given by
\begin{eqnarray}  
& & I_{\alpha 1} \equiv  I +H_{\alpha A}/r_m^\alpha 
+ [r_+^\alpha I_{b+}-r_-^\alpha I_{b-}]/(2 r_m^\alpha),  \nonumber \\
& & H_{\alpha A} \equiv  \left[ \psi_{b+} (z_\alpha)-\psi_{b-} (z_\alpha) 
   - \psi_{\alpha b+} (z_\alpha)+\psi_{\alpha b-} (z_\alpha) \right]/2. \nonumber
\end{eqnarray}
Eqs.~(\ref{accum1}) and (\ref{accum2}) indicate that the spin accumulation is determined by 
the modified current $I_{\alpha 1}$ when the DG term exists. Next, we show that the DG term 
is related with the interface electric field.

{\it Interface electric field.}---
Strictly speaking, the DG terms are numerically determined by solving the Poisson equations and 
the Schr\"{o}dinger equation. However, because many approximations are 
required to obtain interface electronic structure even without spin accumulation~\cite{Stern},
we would like to take the following simple approximation.
From the Schr\"{o}dinger equation
$
d^2 \Psi(z)/dz^2 +[V(z)-E_0] \Psi(z)=0  
$, and the relation  $n=|\Psi(z)|^2$, 
the DG term can be approximately evaluated by
\begin{equation}
\frac{\nabla^2 \sqrt{n}}{\sqrt{n}} \approx \frac{\nabla^2 \Psi(z)}{\Psi(z)} = \frac{2m}{\hbar^2} [V(z)-E_0]. 
\label{VV}
\end{equation}
Then we obtain 
$
I_{\alpha b \pm}^F=\sigma_{\pm} \mathcal{E}^F_\alpha/r_q
$ 
with $\mathcal{E}^F_\alpha=\partial V_F^\alpha /\partial z$
for the FM region
and
$
I_{b \pm}^{N\alpha}=\sigma_{N} \mathcal{E}^N_\alpha/r_q
$ 
with $\mathcal{E}^N_\alpha=\partial V_N^\alpha /\partial z$
for the SC region at the interface of $z_\alpha$.
From the current conservation condition, we have new boundary conditions 
given by
\begin{equation}
S_\alpha [I_{\alpha b+}(z_{\alpha}^-)  +I_{Lb-}(z_{\alpha}^-)    ]
= S_{\rm N} [I_{b+}(z_{\alpha}^+) +I_{b-}(z_{\alpha}^+)], \label{bndy1}
\end{equation}
where $z_{\alpha}$ is the boundary position.
From this condition, we have $S_\alpha \sigma_F \mathcal{E}^F_\alpha=S_{\rm N} \sigma_N \mathcal{E}^N_\alpha$. 
Thus, $\mathcal{E}^N_\alpha$ is determined by $\beta$ of the FM regions.
Then we obtain 
$I_{\alpha 1} = I + 2\sigma_N \mathcal{E}^N_\alpha/r_q$.
Note that, in the present framework, the thickness of the tunneling barrier between FM/SC is neglected 
and the connection between the FC and the SC is described at the boundary point (
$\mu_\pm (z_\alpha^+) -\mu_\pm^\alpha (z_\alpha^-) =r_\pm^\alpha I_\pm (z_\alpha)$).
Therefore, we consider that Eq.(\ref{VV}) can include the main effect of the tunneling phenomena.

\begin{figure}
\centering
\includegraphics[width=6.3cm]{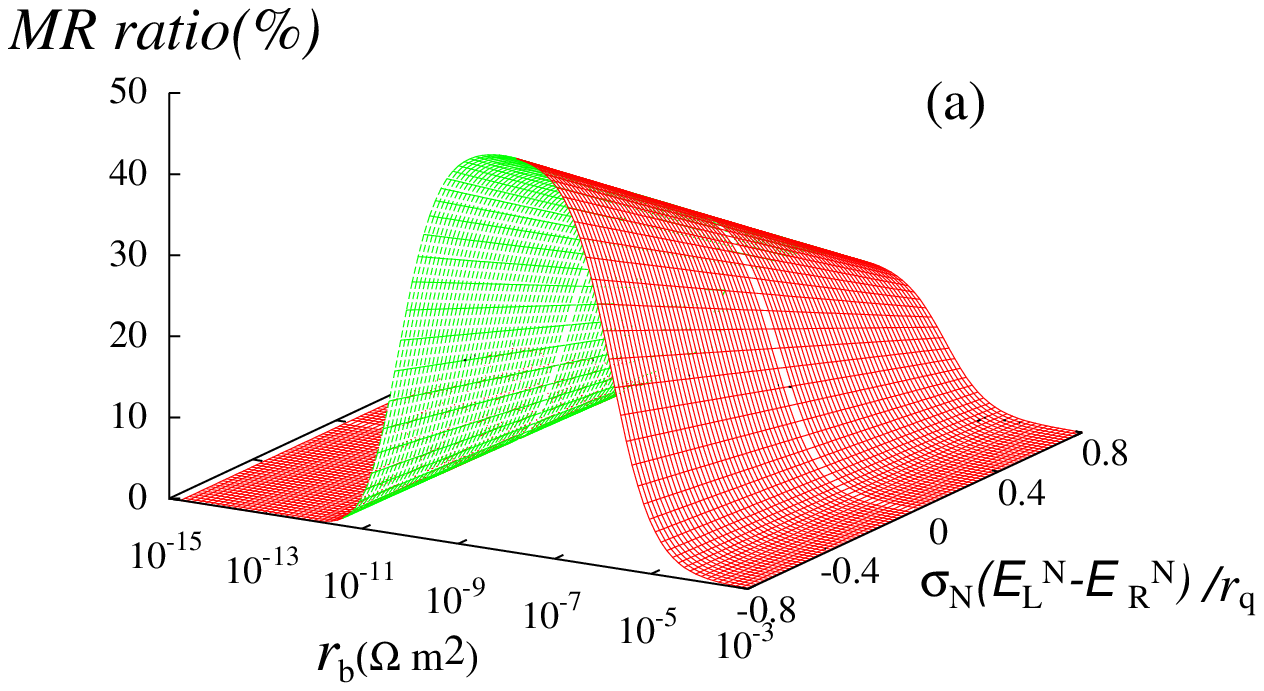}  
\includegraphics[width=6.3cm]{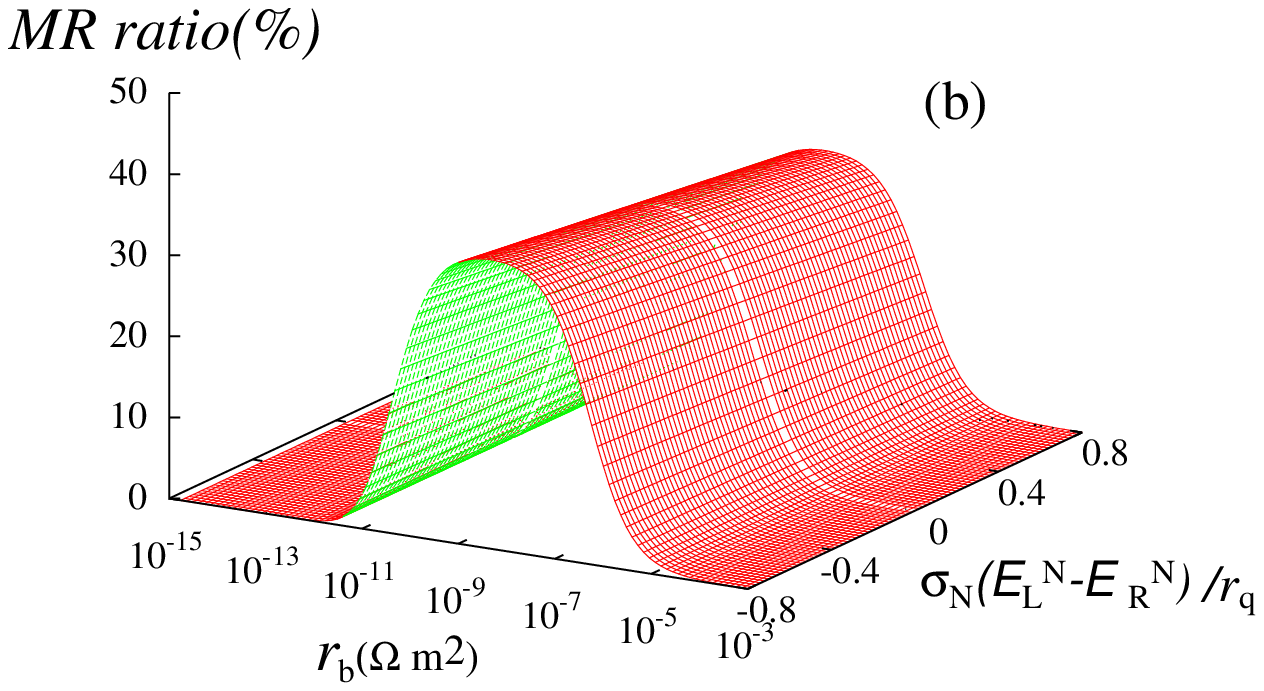}  
\caption{MR ratios of the FM/SC/FM structure (local measurement) as functions of 
the average interface resistance $r_b$ and the asymmetry $y=\sigma_N (\mathcal{E}^N_L-\mathcal{E}^N_R)/r_q$.
The peak corresponds to the conductance matching points.
(a) Area of the left electrode ($S_{\rm L}$) and that of the right area ($S_{\rm R}$) has the relation $S_{\rm L}/S_{\rm R}=1/4$. 
(b) $S_{\rm L}/S_{\rm R}=0.4/0.5$. In this case, we cannot see clear dependence of $y$.
$\gamma_L=\gamma_R=0.5$
$l_N=1~\mu$m, $r_N= 4.0  \times 10^{-9} \Omega~{\rm m}^2$, 
$l_F=5$ nm, $\beta=0.46$, $r_F=4.5\times 10^{-15} ~\Omega~{\rm m}^2$. 
The white line corresponds to the standard theory ($\mathcal{E}^N_L=\mathcal{E}^N_R$).}
\label{MRcal}
\end{figure}

{\it Magnetoresistance.}---
Total resistance through the FM/SC/FM structure is obtained by summation of 
each resistance of the FM and SC elements. 
The resistance difference $\Delta R\equiv r_{\rm AP}-r_{\rm P}$ is given by
\begin{eqnarray}
\lefteqn{ \Delta R = \frac{2 r_N S_{\rm N}}{Ir_F^2\Delta}
\bigl\{ c_F^2 \gamma_L \gamma_R r_b^L r_b^R \mathcal{I}_A 
}\nonumber \\
\!\!\!\!\!&+& \!\! \beta c_Fs_F r_F (\gamma_L r_b^L \mathcal{I}_B  
\!+ \! \gamma_R r_b^R\mathcal{I}_C) 
\!+ \!  4 \beta^2 r_F^2 s_F^2 I S_{\rm N}  \bigr\},
\label{rap}
\end{eqnarray}
where
$c_N\equiv\cosh (t_N/2/l^N)$, $s_N\equiv\sinh (t_N/2/l^N)$, $c_F\equiv\cosh (t_F/2/l^F)$, $s_F\equiv\sinh (t_F/2/l^F)$, and
\begin{eqnarray}
\mathcal{I}_A&=&I_{L1}S_{\rm L} + I_{R1}S_{\rm R}=I[(S_{\rm L}+S_{\rm R})+y(S_{\rm L}-S_{\rm R})],
\nonumber \\
\mathcal{I}_B&=&2I_{L1}S_{\rm L} + I_{R3}S_{\rm R}=I[(2S_{\rm L}+S_{\rm N})+y(2S_{\rm L}-S_{\rm N})],
\nonumber \\
\mathcal{I}_C&=&I_{L3}S_{\rm L} + 2I_{R1}S_{\rm R}=I[(S_{\rm N}+2S_{\rm R})+y(S_{\rm N}-2S_{\rm R})],
\nonumber 
\end{eqnarray}
with $y\equiv (I_{L1}-I_{R1})/(2I)= \sigma_N (\mathcal{E}^N_L(\beta)-\mathcal{E}^N_R(\beta))/(r_qI)$
being the degree of the asymmetry.
The MR ratio is obtained from $\Delta R$ divided by the total resistance of the parallel magnetic 
alignment.
In the impedance matching region $r_N(t_N/l_{sf}^N) \ll r_b \ll r_N(l_{sf}^N/t_N)$~\cite{Fert2001}, we have 
\begin{equation}
MR \approx MR^{(0)}\frac{2(I_{L1}S_L +I_{R1}S_R)}{(S_L+S_R)(I_{L1}+I_{R1})}
\label{MRmain}
\end{equation}
$MR^{(0)}\equiv \frac{\gamma^2}{1-\gamma^2}$ is the MR ratio without the interface effect~\cite{Fert2001}.
We can see that the interface effect modifies the MR ratio by the modified current weighted by area.
When $S_L \approx S_R$, $MR\approx MR^{(0)}$, therefore the asymmetric area is important for increasing 
the MR ratio.
For the symmetric point $y=0$, $\mathcal{I}_A=\mathcal{I}_B=\mathcal{I}_C=I$ and the MR ratio coincides with the conventional formula.
When the electrons flow from the left electrode to the right electrode, 
$\mathcal{E}^N_L <0$ and $\mathcal{E}^N_R =0$(Fig.1), and 
when the electrons flow from the right electrode 
to the left, $\mathcal{E}^N_L =0$ and $\mathcal{E}^N_R <0$. 
Thus, by setting the different interface electric fields depending on the current direction, 
we can obtain a different MR ratio.
These are the main theoretical results of this paper.

Let us confirm these analyses by numerical calculations
assuming $r_p^L=r_p^R$. 
Figure~\ref{MRcal} shows the numerical results of the MR ratio 
as a function of the asymmetry $y$
for (a) $S_{\rm L}/S_{\rm R}=1/4$ and (b) $S_{\rm L}/S_{\rm R}=0.4/0.5$. 
When the asymmetry of the area is large ((a)),  
the effect of the asymmetric local field $y \neq 0$ becomes large.
Moreover, in the region of $y <0$, which 
corresponds to $\mathcal{E}_L^N<\mathcal{E}_R^N$, MR ratio increases.
This can be understood as follows:
the larger depletion region generates larger electric field that accelerates 
electrons resulting in the larger MR ratio.

In Ref.~\cite{tanaJJAP}, we theoretically showed the MR ratio has its maximum at the symmetric structure ($y=0$)
in the range of the standard theory. 
When there is no interface electric field, $I_{L1}S_L=I_{R1}S_R$, then Eq.(\ref{MRmain})
becomes $MR/ MR^{(0)} \approx 2 S_L S_R/(S_L+S_R)^2$.
Thus, the MR ratio has its maximum when $S_L=S_R$.
This means that a new degree of freedom, $i.e.$, local electric field, 
enables the change of the MR ratio.

\begin{figure}
\centering
\includegraphics[width=8.5cm]{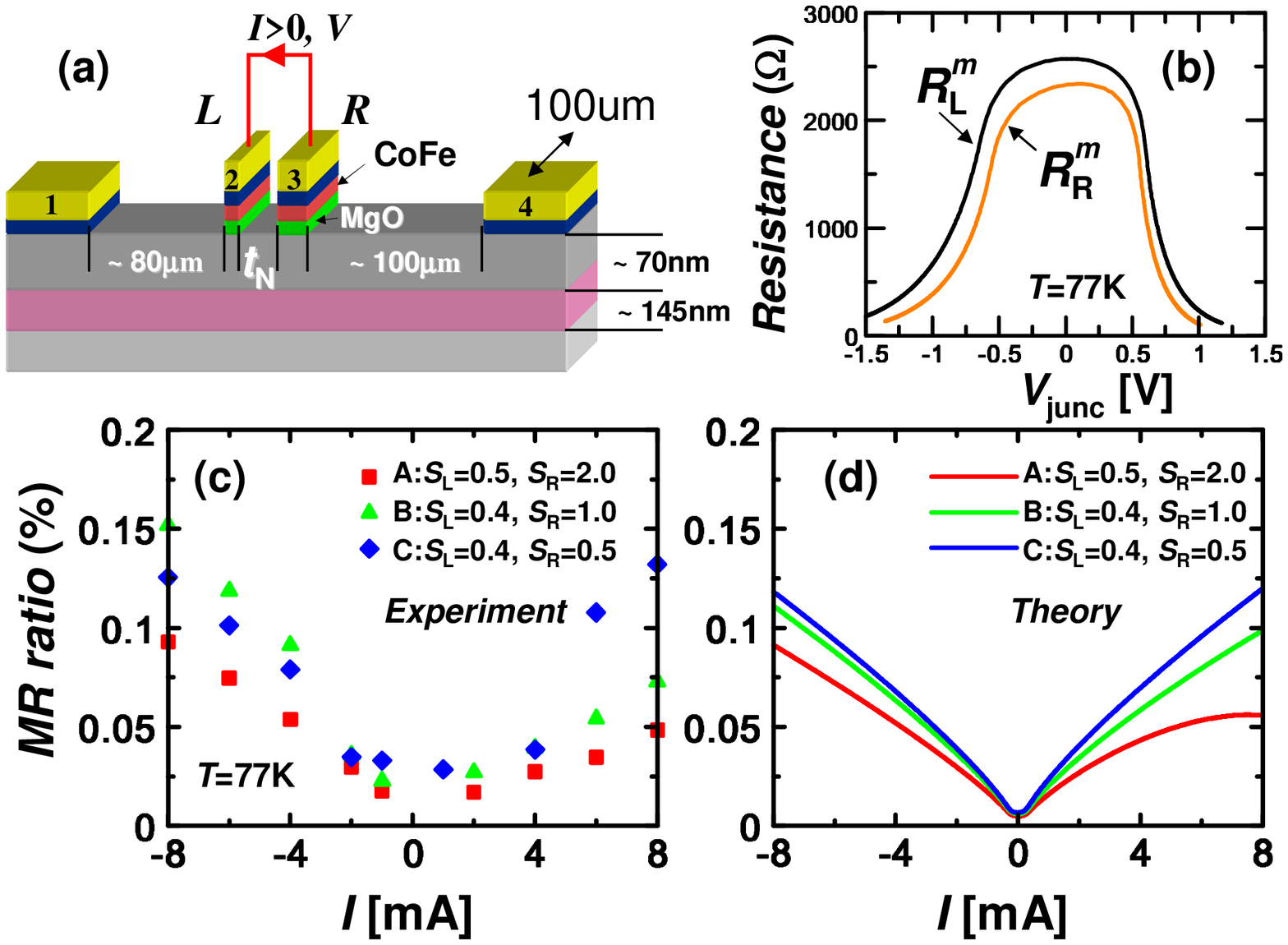}  
\includegraphics[width=6.5cm]{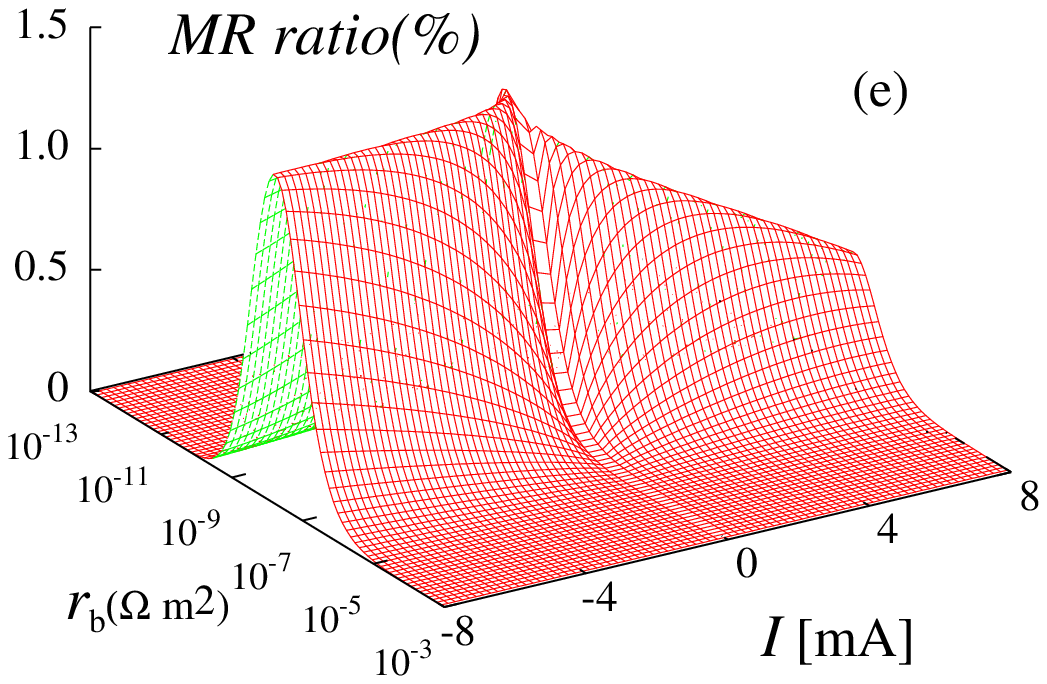}
\caption{(a) Four-terminal devices used in the experiment. 
We prepared three types of devices with different
electrode areas. (b) Measured resistance as a function of junction bias.
The left resistance $R_{\rm L}^{m}$ is measured between terminals 1 and 2. 
The right resistance $R_{\rm R}^{m}$ is measured between terminals 3 and 4.
(c) MR ratio as a function of current through electrodes 2 and 3.
(d) Corresponding theoretical calculation based on Eq.~(\ref{rap}) 
$r_N= 4.5\times 10^{-9}~\Omega~{\rm m}^2$.
$r_F=4.5\times 10^{-15}~\Omega~{\rm m}^2$. 
Other parameters are derived from experiments including the resistance dependence on current of Fig.(b).
$l_N=5.1~\mu$m, $\rho_N=2.09\times 10^{-3} \Omega{\rm cm}$, , 
$l_F=5$ nm. We assume $\gamma_L=\gamma_R=\beta=0.11$ coincide 
with the polarization obtained from the four-terminal Hanle experiments.
(e)Bird-eye view of numerical results.}
\label{Fig_Experiments}
\end{figure}

{\it Experiments.}---
In order to realize the asymmetric electric field, 
areas of two FM electrodes should be different.
We fabricated four-terminal devices for
Hanle-effect measurements, as shown in Fig.~\ref{Fig_Experiments}(a)
for three types of different area configurations:
[A] $S_{\rm L}=50 \mu {\rm m}^2$, $S_{\rm R}=200 \mu {\rm m}^2$ and $t_N=2050$~nm, 
[B] $S_{\rm L}=40 \mu {\rm m}^2$, $S_{\rm R}=100 \mu {\rm m}^2$ and $t_N=1500$~nm, 
[C] $S_{\rm L}=40 \mu {\rm m}^2$, $S_{\rm R}=50 \mu {\rm m}^2$ and $t_N=1250$~nm.  
The CoFe/MgO(1nm) is patterned on a phosphorus-doped 
($\sim 2\times 10^{19}~{\rm cm}^{-3}$) (100) textured Si
of an insulator (SOI) substrate, where ohmic pads consisting
of Au/Ti were formed for all the contacts.
The structures will be discussed in detail elsewhere~\cite{Ishikawa2}.
Figure~\ref{Fig_Experiments}(b) shows the resistance-voltage ($R_{\rm res}$-$V$)
characteristics of the left ($L$) and the right ($R$) tunneling junctions for the device $A$,
which are derived from current-voltage ($I$-$V$) measurements. Here, 
positive bias $V_{\rm junc}>0$ corresponds to a case where electrons flow from SC to FM.
The measured resistance of the $L$ junction, $R_{\rm L}^m$, is always larger than 
that of the right electrode, $R_{\rm R}^m$, because $S_{\rm L}<S_{\rm R}$. 
In addition, it is found that the resistances of $V_{\rm junc}<0$ are 
larger than those of $V_{\rm junc}>0$. 
This means that the depletion layer in SC appears when electrons flow from FM to SC
while there is no depletion layer in SC when electrons from SC to FM.
Hence, the resistance $V_{\rm junc}>0$ can be regarded as the intrinsic junction resistances 
($R_{\rm L}$ and $R_{\rm R}$). \color{black}

Figure ~\ref{Fig_Experiments}(c) shows MR ratio as a function of current $I$ 
through the terminals 2 and 3 (local measurement, see also Fig.~\ref{Fig_Experiments}(a) 
for the current direction). 
In the standard theory, MR ratio is described as a function of current, and therefore, 
hereafter we analyze our experiments as a function of current through terminal 2 and 3.
 We can see that (i) the MR ratio changes 
in proportion to the current (or bias), and (ii) the MR ratio differs depending 
on $I>0$ or $I<0$. Remember that MR ratio of the single junction has a peak 
at the zero bias (zero current)~\cite{Jansen0,Luders}. In addition, 
the MR ratio derived by the standard theory\cite{Fert2001} does not include 
the direction of current. 
These phenomena can be understood by considering the local electric fields as follows: 
$I>0$ ($I<0$) corresponds to a case when $V_{\rm junc} >0$ ($V_{\rm junc} <0$) 
for $L$-junction and $V_{\rm junc} <0$ ($V_{\rm junc} >0$)
for $R$-junction. 
Because of the area difference and Fig.~\ref{Fig_Experiments}(b), 
the depletion region
of the right junction for $I>0$ is smaller than that of the left junction for $I<0$, 
and $|\mathcal{E}_L^N|>|\mathcal{E}_R^N|$ is realized. Then, 
MR ratio for $I<0$ is larger than that for $I>0$.  
These results show a new picture of the spin transport in the local measurement.
The result that this asymmetry is clearest for device $A$ and least for $C$ supports the view
that the asymmetry comes from the strength of the interface electric field.

Next let us quantitatively compare the theory with the experiment. 
In the metal-insulator-semiconductor(MIS) interface,
we have $V_{\rm junc}+V_{\rm FB}+V_{\rm traps}=\mathcal{E}_s d_s+\psi_s$, 
where $V_{\rm FB}$, $V_{\rm traps}$, $\mathcal{E}_s$,$d_s$, and $\psi_s$ are 
the flat-band shift, the voltage fluctuation by trap states, the electric field applied to the junction, 
the thickness of the junction and the surface potential, respectively~\cite{Sze}.
$\mathcal{E}_s$ is estimated by donor density $N_D\sim 2\times 10^{19}~{\rm cm}^{-3}$ and 
temperature $T=77$ K with $d_s=1.5$~nm at low-bias region ($<$ 2~V).
In this region, the number of holes is neglected and 
electrons at $V_{\rm junc}<0$ are in a depletion region with 
$\mathcal{E}_s \approx \sqrt{2eN_D\psi_s/\epsilon_{\rm Si}}(\epsilon_{\rm Si}/\epsilon_{\rm MgO})$.
The coefficient $r_{q}$ is used as a fitting parameter, because 
the interface electric field is not uniform in view of the geometry and the interface roughness.
Concretely we take 
$r_{q}\approx 3.66$ assuming 5~nm depletion layer with the average resistance and current estimated from the experiment.
When the parameters such as Fig.~\ref{Fig_Experiments}(b)
are included, the numerical MR ratio is given in Fig.~\ref{Fig_Experiments}(d).
As can be seen, we can realize the overall tendency of the experiment.
If we apply the experimental parameters to the standard theory, the MR ratio for $I>0$ is larger than that for $I<0$. 
From Fig.3, the difference between the left and the right junction resistances is smaller for $I>0$ than that for $I<0$. 
By considering that the symmetric structure produces larger MR in the standard theory, 
the MR ratio of $I>0$ is larger than that of $I<0$ in the standard theory.

Figure~\ref{Fig_Experiments}(e) shows the MR ratio of devices $A$ as functions 
for the interface resistance $r_b$ and the current. 
The difference from Fig.~\ref{MRcal} is that we use the experimental $R_{\rm res}$-$I$ characteristics (Fig.~\ref{Fig_Experiments})
in which $R_{\rm L}$ and $R_{\rm R}$ increases when $I$ decreases.
This is the reason why MR ratio increases as a function current in Fig.~\ref{Fig_Experiments} (c)
($r_b$ of our experiments is larger than the conductance matching peak). 
The MR ratio of the device $B$ is larger than that of the device $A$ because 
the distance between two electrodes $t_N$ of the former is smaller than that of the latter.
From Fig.~\ref{Fig_Experiments}(e), it is also predicted that the MR ratio decreases as current increases when 
$r_b$ is smaller than its impedance matching point(backside of Fig.~\ref{Fig_Experiments}(e)).

{\it Discussion.}---
Kameno {\it et al.} showed that the spin drift makes lifetime depend on applied voltage~\cite{Yu,Kameno}.
We conducted a lifetime measurement in a three-terminal setup and found there was no bias-voltage dependence on the lifetime~\cite{Ishikawa2}.
This is considered to be because the FM areas of our devices are more than two times larger than those in Ref.~\cite{Kameno}. 
Note that even a small change in the MR ratio requires a larger change of the lifetime $\tau_N$ because $l_N \propto \sqrt{\tau_N}$. 
Thus, the spin drift seems not to be the main cause of the change of the MR ratio in our devices.
Thus, the change of the MR ratio mainly 
comes from the asymmetric electronic structure, including the area difference indicated by Eq.(\ref{rap}).

The effect of trap sites has not yet been explicitly included in the proposed theory. 
Change of lifetime as shown in Ref.~\cite{Tran} might be the direct way 
to include the effect of trap sites. 
The trap sites could also affect the MR ratio through the change of $V_{FB}+V_{\rm traps}$ from which  
$\mathcal{E}_s$ is estimated. 
Here, we treat the coefficient $r_q$ as a fitting parameter, because
the determination of the coefficient $r_q$ requires more detailed experiments.
The estimation of the effect of traps is a subject for future work. 

In summary, we introduced the density-gradient effect into the standard spin diffusion theory, and 
conducted both the local and nonlocal measurements on n-type Si devices.
We showed that the increase/decrease of the MR ratio comes from the asymmetry of  
the two interface electronic structures and explained why the experimentally obtained MR ratio
depends on the current direction.

We thank A. Nishiyama, K. Muraoka, S. Fujita and K. Tatsumura for fruitful discussions.


\end{document}